\documentclass[journal,12pt,onecolumn,draftclsnofoot,]{IEEEtran}

\usepackage[T1]{fontenc}
\usepackage[english]{babel}

\usepackage{algorithm}
\usepackage{algpseudocode} 
\usepackage{acronym}
\usepackage{graphicx}
\usepackage{epstopdf}
\usepackage{amsthm}
\usepackage{amsmath}
\usepackage{amssymb}
\usepackage{tabularx}
\usepackage[font=scriptsize]{subfig}
\usepackage{xcolor}
\usepackage[font=scriptsize,labelfont=bf]{caption}

{\left\lbrace\begin{array}{@{}l@{}}}%
{\end{array}\right.}

\begin{document}
	
\def\OurTitle{Online and Global Network Optimization\\
	{\LARGE Towards the Next-Generation of Routing Platforms}}

\def\OurAuthors{{J\'er\'emie Leguay, Moez Draief, Symeon Chouvardas, Stefano Paris,\\ Georgios S. Paschos, Lorenzo Maggi, Meiyu Qi\thanks{Email: firstname.lastname@huawei.com}}}

\title{\OurTitle}
\author{\IEEEauthorblockN{\OurAuthors\\}
	\IEEEauthorblockA{\vspace{0.3cm}{\normalsize Mathematical and Algorithmic Science Lab,\\\vspace{-0.3cm} France Research Center, Huawei Technologies.}}
}
\maketitle

\begin{abstract}
The computation power of SDN controllers fosters the development of a new generation of control plane that uses compute-intensive operations to automate and optimize the network configuration across layers. From now on, cutting-edge optimization and machine learning algorithms can be used to control networks in real-time. This formidable opportunity transforms the way routing systems should be conceived and designed. This paper presents a candidate architecture for the next generation of routing platforms built on three main pillars for admission control, re-routing and monitoring that would have not been possible in legacy control planes.
\end{abstract}


\section{Introduction}

Software-Defined Networking (SDN) technologies~\cite{kreutz2015software} are radically transforming the network architecture of data centers, network overlays, and carrier networks. They provide programmable data planes that can be configured from a remote controller platform. SDN controllers are implemented on top of commodity servers which provide a tremendous computational power compared to legacy network devices. They also use modern distributed computing technologies (e.g., loosely consistent databases, parallel computing, consensus algorithms, transactional updates) to gather and keep a global view of the network status in real-time and to push consistent configuration updates to the network equipment.

While initial SDN deployments focused mainly on automated provisioning of tunnels and virtual networks, essentially virtualizing known concepts, it becomes clear today that SDN presents an opportunity for a paradigm shift in network optimization. Departing from the classical autonomic routing systems we migrate to a new centralized routing paradigm, where a powerful controller is aware of the static parameters and time-varying conditions, and centrally orchestrates the entire network reaching unprecedented levels of efficiency. 
Indeed, the necessary ingredients are in place: (i) the power of modern SDN controllers, and (ii) the recent advances in optimization and machine learning theories can be married to produce the desired outcome. 
This proof of concept was recently demonstrated in large scale datacenter interconnects with the B4 controller~\cite{jain2013b4}.    



This paper presents a candidate architecture for compute-intensive control planes where new types of algorithms can be deployed for online and global network optimization. 
It fully exploits the processing power and the central position of the controller to globally optimize routing and admission control using efficient mathematical tools ranging from online algorithms, combinatorial optimization, control theory and machine learning. It improves the accuracy of network monitoring at low overhead costs by using data completion techniques and online learning algorithms to better coordinate the data collection effort. Finally, it tightens the link between monitoring and routing which can now co-exist on the same platform. 

Our candidate architecture for the next-generation of routing platforms relies on three fundamental building blocks:

\textit{Admission control with experts.} 
The centralized nature of SDN architectures unleashes the potential of using online algorithms which have been initially defined for online covering and packing problems with provable guarantees compared to an offline optimal. Due to the unpredictable nature of network traffic, it is difficult to choose a priori which online algorithm should be used.
We thus perform admission control with expert advice. This setting is classical in machine learning, and it is commonly called  boosting algorithms. Past decisions and past rewards obtained by the different algorithms are used to track and follow the advice of the best algorithm in hindsight.

\textit{An iterative routing solver.} To maintain the best flow configuration in the network, we propose to use an iterative routing solver which outputs a sequence of feasible solutions. As the routing solver may work on multi-commodity problems with millions of variables and constraints in large networks, the convergence to the optimal can take a few iterations. To avoid pushing the solution of each iteration into network equipments, we propose a control framework that operates on top of the routing solver and decides which feasible configuration to apply. Given a  reconfiguration budget, it aims at minimizing the flow allocation cost.

\textit{Parsimonious monitoring.} A severe complication that arises in network monitoring is the incredible amount of measurements collected. This \textit{big data} includes important information for network control, but due to its immense size is difficult to handle. Although the computational power available at SDN controllers is an helpful ally, intelligent data handling is also necessary. In fact, recently developed algorithmic tools from machine learning, graph theory, and signal processing can be efficiently combined to enable the SDN controller to have an accurate view of the time-varying network conditions and their evolution. We show how to combine 1) perform parsimonious and adaptive network measurements, 2) recover missing information and infer global traffic properties from the measurements, and 3) predict future variations in network traffic using spatial and temporal correlations.


The rest of this article is organized as follows. After describing the overall architecture of the routing platform., we present the core elements for admission control, re-routing and real-time monitoring. For each of them we discuss their operational benefits and future directions.

\section{Architecture}

\begin{figure*}[!t]
	\centering
	\includegraphics[width=0.7\linewidth]{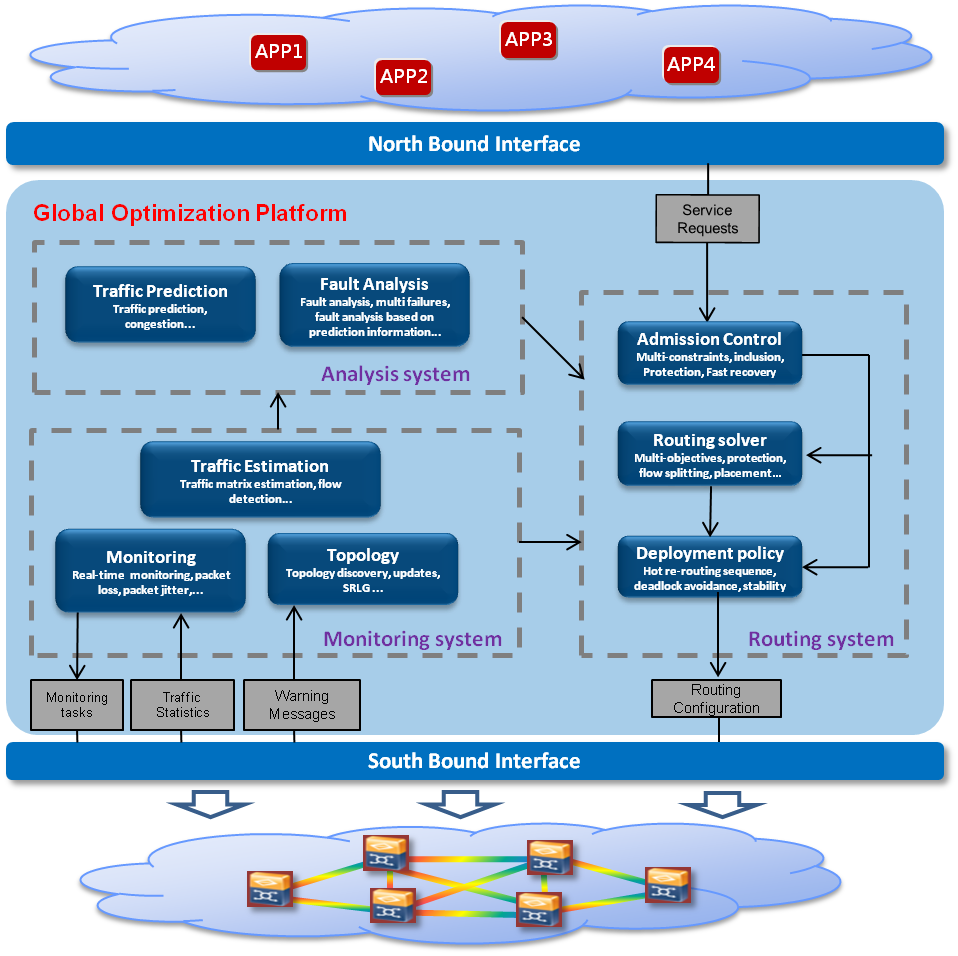}
	\caption{Global Routing Optimization Platform with three main pillars for admission control, re-routing and monitoring.}
	\label{fig:architecture}
\end{figure*}

We have in mind a classical SDN controller architecture depicted in Fig.~\ref{fig:architecture}. The controller is connected through north bound interface to the application layer enabling a wealth of applications like bandwidth on demand, load balancing, etc, while the decided routing commands are dispatched to the network devices using SDN protocols such as OpenFlow via the south bound interface. The latter is also used to collect measurements. This work is primarily concerned with highlighting the architecture of the SDN controller. In Fig.~\ref{fig:architecture}  we depict the three fundamental building blocks of our architecture, namely 1) the monitoring system, 2) the analysis system, and 3) the routing system. Below we explain in detail each system operation and functionality.

SDN controllers classically receive service requests from applications and send routing commands to programmable switches. At the same time, they monitor the network status to quickly react to system changes.
Although it might be tempting to reuse control plane algorithms from legacy routing software, their formidable computation and strategic location spur the development of a new generation of control plane that uses compute-intensive operations to automate and optimize the network in real-time. This routing platform can embed a set of complementary optimization and machines learning algorithms which seat together and tightly cooperate. Fig.~\ref{fig:architecture} presents a candidate architecture with three main sub-systems for routing, monitoring and analysis. 

\textbf{Routing system.}
When connection requests arrive at ingress nodes or when a group of demands have to be rerouted after a failure, the controller has to find a set of feasible paths satisfying multiple constraints (e.g., network capacity, QoS, protection). In addition, smart decisions must be taken to accept or reject service request to optimize long term objectives such as throughput maximization or load balancing.  Although this \textit{Fast Connection Setup} problem refers to a single commodity flow problems, it is still quite complicated to solve. Hence at this stage, the goal is not to optimize the network, but rather to quickly find a good path allocation.  

The network configuration obtained as a result of fast connection setups and of flow departures poses significant concerns on the evolution over time of the global objective function (e.g., routing cost). Therefore, periodic or event-based reconfiguration of the overall flow allocation can help reduce the optimality gap. We call this mechanism \textit{Garbage collection of network resources} since it mirrors the way Java reorganizes the memory. For this, an iterative routing solver works on an evolving instance of the routing problem to provide a sequence of better configurations. 
A deployment policy, which works on top of the iterative solver, decides 
which configuration should be applied, taking into account a fixed reconfiguration budget.

\textbf{Monitoring system.} The main objective it to feed the routing system with the most accurate view of the network in terms of traffic and topology. The central position of the controller allows to implement completely new monitoring systems compared to legacy architecture where equipments were statically configured to report periodic traffic samples and statistics. The data collection effort can from now on be driven by the controller to avoid redundant and low value measurements. Traffic sampling can be adapted in real time to maximize the accuracy of traffic estimation algorithms while keeping the overhead low or below a given budget. It can be used~\cite{callado2009survey} to better classify flows in real time (e.g., heavy hitters identification, DDoS detection) and to guide the routing system for the provisioning of custom flow configurations. 

\textbf{Analysis system.} On top of the monitoring system and at a larger time scale, SDN controllers can embed traffic and fault prediction algorithms that can spare the routing system from taking myopic decisions. They can exploit this estimation and its uncertainty to optimize network resources in a robust fashion~\cite{bauschert2014network}.
\section{Admission Control with Experts}

A key task of the SDN controller is the Admission Control (AC) of incoming connection requests, in an online fashion. Its goal is to gracefully manage service requests when the network becomes highly utilized, as new incoming requests arrive. Non-myopic accept or reject decisions have to be made to maximize a given profit, such as the total accepted throughput, the financial revenue generated or the quality of service experienced by users.

The existing admission control implementations are in their majority threshold-based. They use max-, min-, exclusive- and non-exclusive-limits on resource portions that the network operator can define for different classes of flows. At any time, routers inspect the sum of exclusive and non-exclusive portions and perform a threshold decision. If the sum is higher than a certain (engineered) percentage of the network capacity, then new connections are rejected. The main problem here lies in defining the threshold in a dynamic fashion, as the optimal configuration depends on the network traffic conditions, which fluctuate over time.
Several proposals on how to optimize the configuration of thresholds can be found in the literature. Worst-case strategies have been put forward for the sake of network planning.
They are based on sophisticated stochastic models using queuing theory and require some \textit{a priori} information on the stochastic model for the traffic evolution. Such information is typically extracted offline with the aid of long-term traffic estimators available at the controller. 

Admission Control can be formulated as an online packing problem, as the goal for instance is to maximize the number of accepted requests subject to some constraints. The challenge in this context comes from the online nature of the optimization problem. New variables and additional constraints are revealed sequentially, as soon as an arrival or departure of a flow occurs in the system. The theory on online algorithms has evolved significantly in the last decade for this type of problems. Algorithms with theoretic guarantees over the offline optimal, which has full information on the future state of the system, have been proposed for online packing problems.These guarantees hold only when consistent decisions are taken by all devices within the network. The centralized nature of SDN lies the ground to apply online algorithms for admission control~\cite{LeguayNoms16}.

\begin{figure*}[!t]
	\centering
	\includegraphics[width=0.7\linewidth]{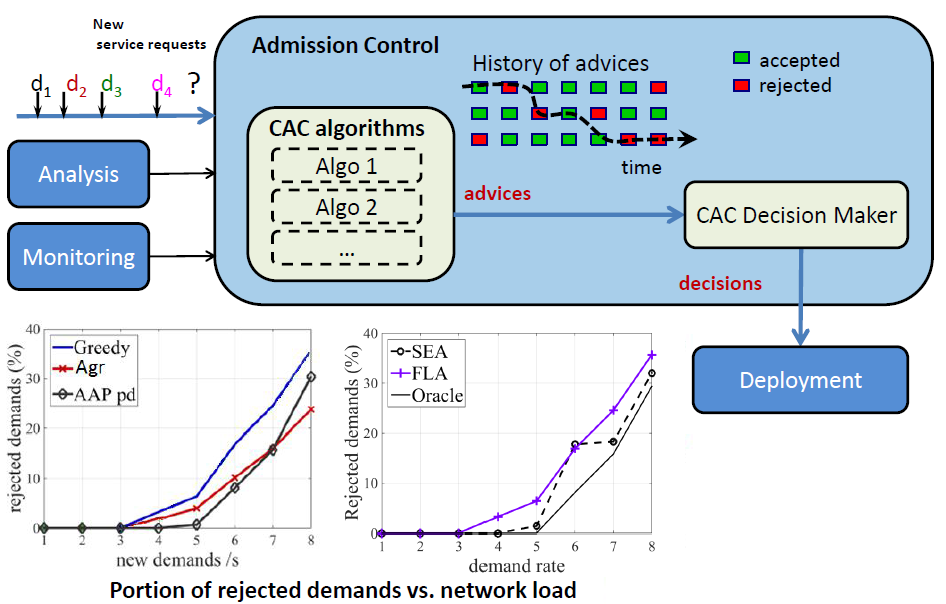}
	\caption{Admission control module with expert advice from Greedy, Agrawal and AAP-PD (Primal Dual). Performance evaluation results on GEANT network (see~\cite{LeguayNoms16} for more details).}
	\label{fig:experts}
\end{figure*}

\subsection{Competitive online algorithms}

Motivated by the considerations above, we first review and adapt, well-known and recent algorithms from the online literature to the admission control problem in SDN. We then test their performance under different traffic conditions to understand and highlight the strengths and weaknesses of each of them. Traditionally, online algorithms for admission control fall into two main categories: i) worst-case and ii) average-case. 

Worst-case algorithms are characterized by max-min performance guarantees under specific worst-case scenarios where a malicious adversary chooses the worst possible sequence of connection requests. Due to their conservative nature, they generally underperform under more standard traffic conditions. One of the first online algorithms for admission control has been AAP. The algorithm is log competitive, meaning that it cannot reject O(log n) times more requests than the offline optimal (n being the number of nodes. Buchbinder et al. proposed in~\cite{BuchbinderN06} a primal dual framework to derive algorithms for online packing and covering problems with performance guarantees in the worst-case scenario. Such framework developed the theory behind the initial intuition of AAP.

On the other hand, ii) average-case algorithms show high expected performance over random traffic conditions, but cannot guarantee good performance in specific adversarial scenarios. The Primal-Beats-Dual (PBD) algorithm has been introduced by Kesselheim et al. in~\cite{Kesselheim2014} . It computes the optimal (fractional) solution of the relaxed Linear Program (LP), by considering all the past requests and scaling the capacity of the graph. It then attempts to route the request over a path randomly selected, with a probability proportional to the value of the computed fractional solution. PBD suffers from computation complexity. Agrawal et al.~\cite{Agrawal2015} have proposed a fast algorithm with multiplicative updates to solve this issue. It applies to i.i.d. and random order inputs. It implements an efficient stochastic gradient descent that we adapted for our admission control problem.

\subsection{Using expert advices}

As observed in practice and highlighted on Fig.~\ref{fig:Monitoring} which plots a performance evaluation on GEANT, worst-case and average-case online algorithms for admission control are better than \textit{Greedy}, the naive strategy which accepts every demand until bottlenecks appears. However, there is no algorithm outperforming all the remaining ones under all traffic conditions. Luckily, modern SDN control platforms, which are running on top of commodity servers and are built upon cutting-edge distributed computing technologies, enable the parallel execution of different algorithms to solve a single decision problem. Such algorithmic architecture, called boosting or prediction with expert advice setting, is commonly used in machine learning. It executes all the algorithms in parallel and attempts to track the best one in an online fashion. 

The bulk of the literature on experts focuses on proving theoretical performance bounds in the basic non-reactive scenario where the action taken by the decision maker does not affect the state of the system, which is definitely not the case for admission control. We identified that the Strategic Expert meta-Algorithm (SEA)~\cite{megiddo2003combine} applies to out reactive setting. Under some stationary conditions on the system, SEA is proven to perform at least as well as the oracle which steadily selects the algorithm with best average performance. As shown on Fig.~\ref{fig:Monitoring}, this approach achieves very good performance in practice. In the worst case, SEA performs as Follow-the-Leader meta-Algorithm (FLA) which alway picks the most instantaneous rewarding decision (i.e., Greedy). However, we believe there is room to improve these encouraging results by better combining the advices and integrating traffic predictions. 

The advent of Software Defined Networks (SDN) advocates a paradigm shift in the implementation of admission control (AC) protocols for at least two main reasons. Firstly, SDN control is centralized, which finally allows to utilize online AC algorithms developed for online packing and covering problems. Secondly, we can now afford to run several online admission control algorithms in parallel and integrate real-time traffic predictions in our decision making.

\section{Iterative Routing Solver}

Traditionally, routers collaboratively share topological information and establish end-to-end paths using traffic engineering protocols such as OSPF and MPLS with TE extensions~\cite{Wang08}.
However, TE solutions that rely on hot standby mechanisms to quickly handle topological changes (in case of link failure) suffer from possibly long recovering times, while other mechanisms at IP level, more reactive, do not offer a fine-tuned control of paths and may lead to sub-optimal flow allocations.
Hence these approaches fall short of optimizing network utilization while meeting QoS requirements and guaranteeing resilience at the same time.
Conversely, SDN architectures unleash the potential to reconfigure in real-time custom paths for each flow, so that network resources are optimized and performance requirements are met (in QoS, resilience). 

\begin{figure*}[!t]
	\centering
	\includegraphics[width=0.7\linewidth]{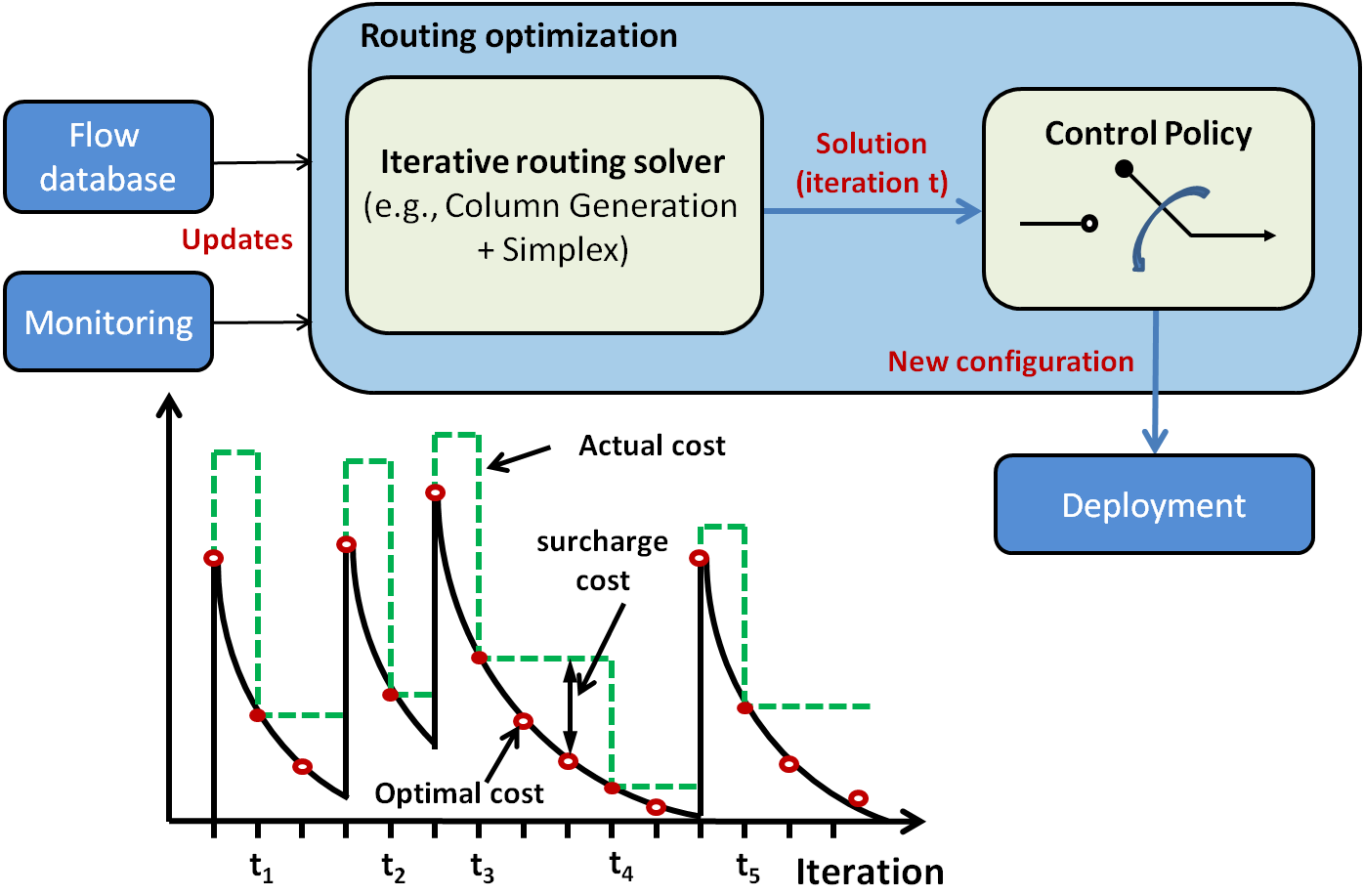}
	\caption{Routing optimization module based on an iterative routing solver and a policy to control its output (top). Evolution of the optimality gap obtained using the iterative  routing solver (red points) and the control policy (dashed green curve). The black curve is the interpolation of the points computed by	the solver at each iteration. Solid dots show where the solution computed by the solver is applied.}
	\label{fig:Policy}
\end{figure*}

To maintain the best network flow configuration, the SDN controller has to solve variants of the classical Multi-Commodity Flow (MCF) problem
in real-time, which may involve millions of variables and
constraints in large networks. 
Several and different constraints can be imposed on each commodity such as node or edge inclusion, protection mechanisms with a maximally disjoint backups paths, or a maximum end-to-end delay. In addition, some flows may receive a special treatment such as elephant flows which must be split over multiple paths while only a few split configurations are possible due equipment limitations (e.g., hash functions, group tables).
As the overall problem instance itself
evolves over time due to time-varying demands or topological failures, the SDN controller solves a sequence of routing problems and needs to
constantly reconsider the flow configuration. Finally, to satisfy
application requirements, the controller needs to solve these
problems under tight timing constraints.
Due to the large problem size,
the routing solver has to iterate to deliver a sequence of feasible configurations. The sequence of sub-optimal network configurations obtained as a result of the iterative execution of the solver typically exhibits diminishing returns. 

At each time slot the SDN controller has the option to reconfigure the network flows as per the current computed solution. However, flow reconfigurations degrade QoS and introduce inertia into the system, hence we may avoid them and neglect to apply the most recent computed configuration. Ideally, the objective function of the problem includes a cost that models the distance between two resource allocations (e.g., the number of edge or flow changes) so that the iterative solver finds a good trade-off between the allocation and the reconfiguration costs. On the other hand, using the existing configuration may
quickly become costly since the new flows that are coming, or the ones that need to be repaired after a network failure, are configured with quick suboptimal routing decisions.

We propose a control framework that operates on top of the iterative SDN routing solver to limit the number of network reconfigurations and to reduce at the same time the flow allocation cost. The control policy applies to switches the solution of the SDN routing solver only when the benefit is larger than a threshold, which is dynamically and optimally updated over time. At any time instant such a dynamic threshold combines the gap between the current and optimal configurations and the number of reconfigurations applied so far. Therefore the control policy strives to balance the surcharge and the reconfiguration rate.

\begin{figure}[!t]
	\centering
	\includegraphics[width=0.6\linewidth]{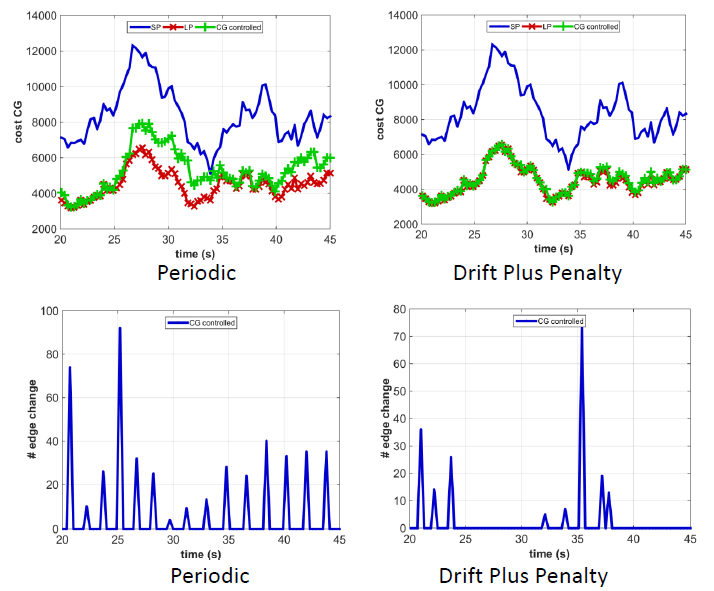}
	\caption{Performance comparison on GEANT in term of routing cost for Shortest Path (SP), the offline optimal (LP), and the policies which limit network reconfigurations (top). Number of routing entry changes at each  reconfiguration with the two policies (bottom).}
	\label{fig:Policy}
\end{figure}

The goal of minimizing the number of flow reconfigurations
clearly conflicts with the goal of minimizing the average optimality gap. It is possible to apply the best found solution at a frequency less than one, which will incur less reconfigurations but it will increase the optimality gap.
We recently proposes~\cite{ParisInfocom16} a drift-plus-penalty policy using a virtual queue as a fundamental tool, which is essentially a counter on which a dynamic threshold is applied.

In order to showcase the validity of our approach, we implemented around an iterative routing solver the control framework with the drift-plus penalty (DP) control policy. As the goal is to stay below a target reconfiguration rate while minimizing the routing cost, we compared the performance of our policy against a Periodic Policy (PP)  that selects reconfiguration points periodically.

Numerical results obtained in real-life network conditions show that SDN controllers can effectively track the evolution of traffic, thus opening a new era for the dynamic control of online routing solvers. Fig.~\ref{fig:Policy} presents a performance comparison between a greedy Shortest Path (SP) where the allocation is never reconfigured, the offline solution of the Linear Program (LP), and the online policies with a limited reconfiguration budget. We can clearly see that DP drastically limits reconfigurations while minimizing the allocation cost.

This framework can be extended towards a number of interesting directions. First, the development of scalable multi-commodity flow solvers that could work on large evolving problems, without the need to converge to the optimal configuration before the system changes. Second, the integration of traffic predictions to avoid myopic decisions with regards to the evolution of traffic so that reconfigurations could be even further limited. 
Third, other metrics that model the distance between two configurations in terms of equipment and flow changes can be designed to be included in the objective function.
Finally, traffic and network uncertainty can be considered~\cite{yang2014traffic} using stochastic and robust optimization techniques.

\section{Parsimonious Monitoring}
The inherent centralized nature of SDN controllers requires a non-negligible amount of signaling between SDN switches, monitoring the traffic over links, and the SDN controller, exploiting such information to take routing decisions.
When both the network size and the frequency at which new requests arrive are large, the need to reduce the signaling traffic arises. Hence, parsimonious collection of statistics, e.g., flow statistics, link loads, just to name a few, becomes essential.  For this reason,
we propose to undersample the measurement collection 
 and to exploit techniques which spring from the compressed sensing algorithmic family, such as matrix completion techniques, to estimate the missing information. The rationale behind this approach is that the observed traffic exhibits inherent time and space dependencies and that one can take advantage of such dependencies to collect partial information on the traffic characteristics and yet be able to complete the missing information. 
 
 

\begin{figure*}[!t]
	\centering
	\includegraphics[width=0.7\linewidth]{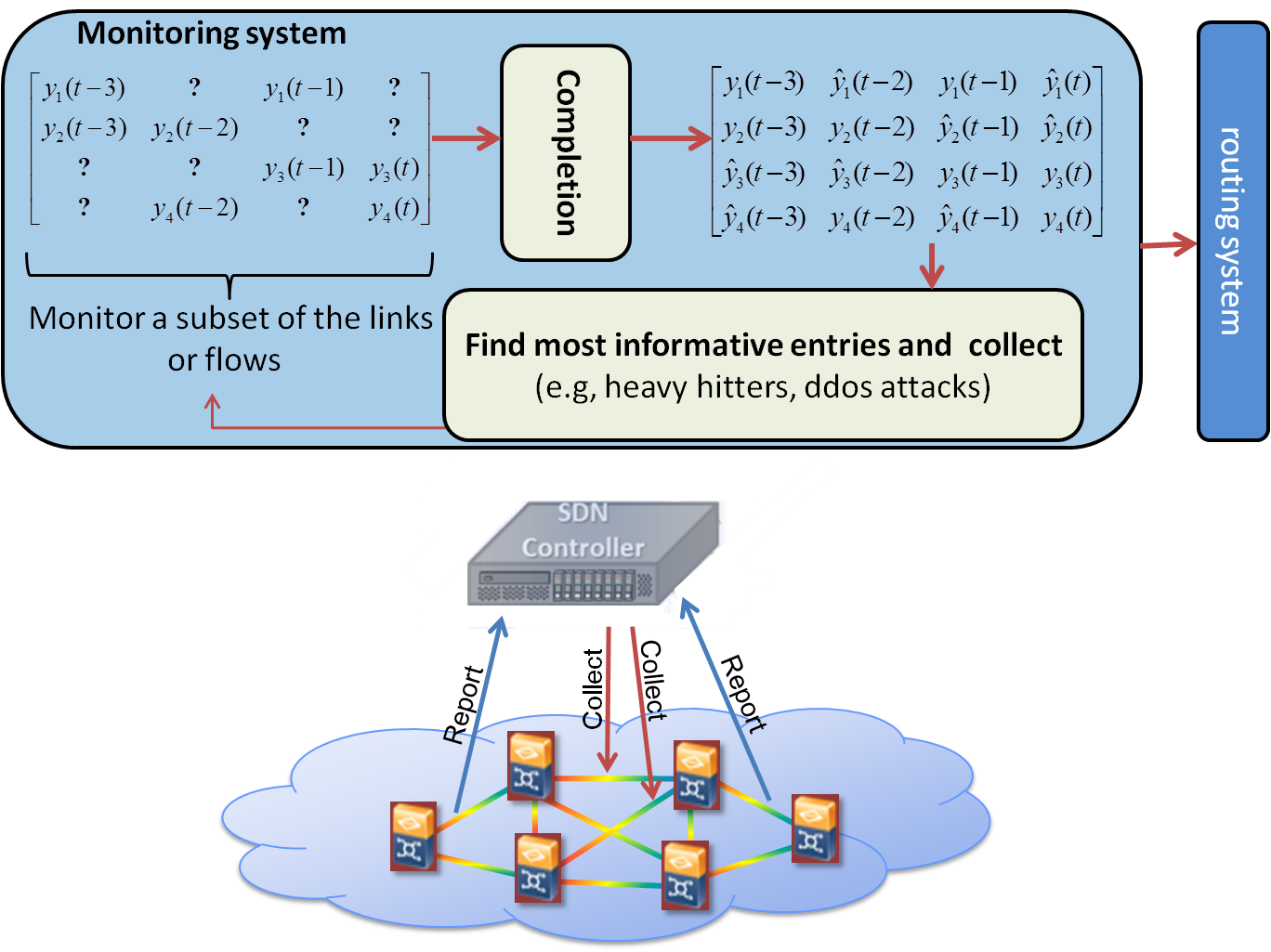}
	\caption{Parsimonious monitoring process driven by the controller.}
	\label{fig:Monitoring}
\end{figure*}

\begin{figure}[!b]
	\centering
	\subfloat[Rejected demands]
	{
		\includegraphics[width=0.3\linewidth]{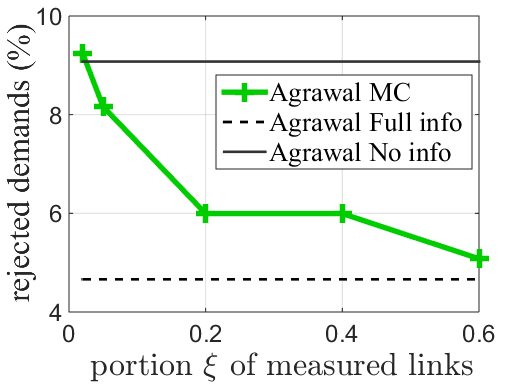}
	}
	\subfloat[Monitoring Error]
	{
		\includegraphics[width=0.3\linewidth]{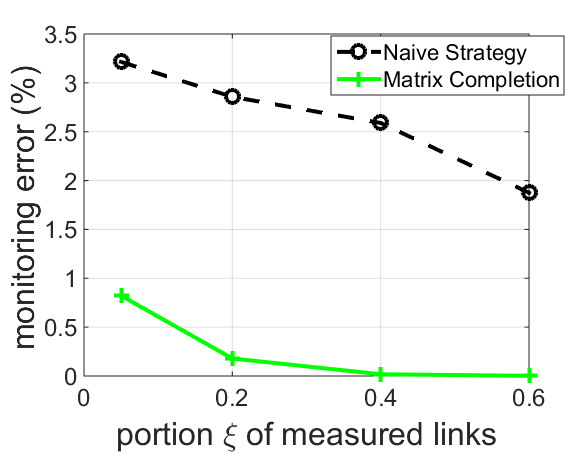}
		
	}
	\caption{Left: percentage of rejected demand vs. measurement
		portion $\xi$ under Random traffic conditions, for Agrawal's AC
		algorithm with matrix completion (Agrawal MC), with full information (Agrawal Full info), and without information about the presence of best-effort traffic (Agrawal No info). Right: Total estimation error over all matrix elements for SVT matrix completion and the naive strategy, vs. the portion $\xi$ of measured links.}
	\label{fig:monitoring-results}
\end{figure}

We propose a framework to parsimoniously and intelligently collect information on link loads. We assume that one has only access to the data from a few links at any given time instant (see bottom of Fig.~\ref{fig:Monitoring}) and that we wish to recover all the link loads at every instant to inform the routing system. In this  context, there are clear spatial and temporal dependencies in the traffic patterns that can be exploited. More specifically, these dependencies are manifested in the low dimensionality of the observed data. The space-time traffic matrix capturing the link loads evolution over time is inherently low rank. To complete the missing data, we use the matrix with the \textit{lowest} rank consistent with the observed data. In Fig.~\ref{fig:monitoring-results}.a, we report the approximation error of the aforementioned matrix completion algorithm and compare it to naive approach that completes the missing load for a given link using the most recent available measurement for that link when we increase the number of observed links per time-epoch. 

We also use these link-load estimates to run the admission control algorithm in a scenario where guaranteed services coexist with best effort traffic which fluctuates over time, see Fig.~\ref{fig:monitoring-results}.b. More precisely, online routing algorithms ideally need a pervasive and seamless
knowledge of the residual capacity on each link. For this reason, we
propose a method to parsimoniously monitor the best-effort traffic on subsets of links and then estimate the missing entries via matrix completion techniques, e.g., Singular Value Thresholding (SVT). We show that the introduction of traffic estimation techniques is beneficial both in terms
of reduction of rejected demands. We evaluated the effectiveness of all
the proposed solutions on the GEANT topology.

The above paragraph assumes that we only have access to link loads and leverage these direct measurements to complete the missing information. In many realistic SDN settings, the controller can retrieve several counters about flows. In OpenFlow for instance, a counter is maintained for every single rules in flow tables. Each entry can correspond to a specific flow by matching on the traditional 5-tuple or to a given flow aggregate. Based on this information, the goal is for example to recover the complete origin destination traffic matrices over time. A simple way of modeling such a problem is to see it as a linear system where we the unknown vector is the per flow traffic and what we observe the vector of aggregate flows as the (matrix) product of the aggregation matrix and the individual flow vector.  It is worth noting that the number of observed aggregate flows is generally much smaller than the actual number of flows in the system. Indeed, rules are stored in TCAM of hardware switches. This type of memory can perform high-speed packet lookups
but are extremely expensive and thus limited in space.

In a recent paper, a system called iSTAMP \cite{malboubi2014intelligent}, such an under-determined system is solved by adding a sparsity constraint to the flow vector (analogous to low-rank constraints for link-load matrices). The rationale behind this assumption is that a small number of flows explain the observed traffic. The iStamp proposal goes a step further and extends the setting to adaptive measurements aiming at monitoring the most informative flows to complete the information gleaned from aggregate measurements. In particular, given the opportunity to directly observe a few individual flows, which flows should one measure to \textit{optimally} improve the error in solving the above under-determined linear sysem. To this end, a multi-armed bandit approach is proposed to select the best (called most rewarding in \cite{malboubi2014intelligent}) such individual flows in terms of their estimated size. This approach is shown to have a very good performance in the context of heavy-hitters identification.

Another popular approach to monitor networks is to sample packets  \cite{moshref2014dream,yu2013software}. Due to the vast amount of information sampling every packet and tracking each flow of the network is not feasible. 
To circumvent this, modern systems implement efficient mechanisms for picking the relevant packets to decide which specific flow aggregates to measure. They rely on hash functions (mapping the data into an array of buckets) to build a compact summary of the stream of packets. These summaries and associated estimated counters are then exploited to uncover traffic statistical patterns such heavy hitters or potential sources of denial of service attacks among other traffic patterns.

One of the aims of this paper is to provide some preliminary
ideas, and hopefully to spur new theoretical research, on
the new and interesting field on parsimonious monitoring for
admission control. In particular, we envision to explore the field 
of \textit{adaptive sampling} through which we will identify and then 
sample the most informative
pieces of information so that to fulfill our monitoring task.
 Furthermore, we envision to explore the very interesting field
 of monitoring of data streams via sketching 
 and to reveal how such techniques could be combined with the compressed sensing rationale.

\section{Perspectives}

As seen in this paper, the next-generation of routing systems for online and global network optimization is likely to embed computer-intensive machine learning and optimization algorithms. Beyond admission control, re-routing and monitoring it will also embed algorithms for fault prediction and analysis to drive real-time decisions in a robust fashion.

The tremendous computation power of SDN controllers does not mean that a single entity will solve everything in real-time. Routing platforms will still follow multi-domain and hierarchical organizations which can help to break down large network optimization problems. In this context, all the algorithms will have to be extended to interact with their peers or superiors to reach a global optimization objective.

\bibliography{IEEEfull,biblio}

\begin{thebibliography}{10}
\providecommand{\url}[1]{#1}
\csname url@samestyle\endcsname
\providecommand{\newblock}{\relax}
\providecommand{\bibinfo}[2]{#2}
\providecommand{\BIBentrySTDinterwordspacing}{\spaceskip=0pt\relax}
\providecommand{\BIBentryALTinterwordstretchfactor}{4}
\providecommand{\BIBentryALTinterwordspacing}{\spaceskip=\fontdimen2\font plus
\BIBentryALTinterwordstretchfactor\fontdimen3\font minus
  \fontdimen4\font\relax}
\providecommand{\BIBforeignlanguage}[2]{{%
\expandafter\ifx\csname l@#1\endcsname\relax
\typeout{** WARNING: IEEEtran.bst: No hyphenation pattern has been}%
\typeout{** loaded for the language `#1'. Using the pattern for}%
\typeout{** the default language instead.}%
\else
\language=\csname l@#1\endcsname
\fi
#2}}
\providecommand{\BIBdecl}{\relax}
\BIBdecl

\bibitem{kreutz2015software}
D.~Kreutz, F.~M. Ramos, P.~Esteves~Verissimo, C.~Esteve~Rothenberg,
  S.~Azodolmolky, and S.~Uhlig, ``Software-defined networking: A comprehensive
  survey,'' \emph{Proc. IEEE}, vol. 103, no.~1, pp. 14--76, 2015.

\bibitem{jain2013b4}
S.~Jain, A.~Kumar, S.~Mandal, J.~Ong, L.~Poutievski, A.~Singh, S.~Venkata,
  J.~Wanderer, J.~Zhou, M.~Zhu \emph{et~al.}, ``{B4: Experience with a
  globally-deployed software defined WAN},'' in \emph{{Proc. ACM SIGCOMM}},
  2013.

\bibitem{callado2009survey}
A.~Callado, C.~Kamienski, G.~Szab{\'o}, B.~P. Ger{\"o}, J.~Kelner,
  S.~Fernandes, and D.~Sadok, ``A survey on internet traffic identification,''
  \emph{Communications Surveys \& Tutorials, IEEE}, vol.~11, no.~3, pp. 37--52,
  2009.

\bibitem{bauschert2014network}
T.~Bauschert, C.~Busing, F.~D'Andreagiovanni, A.~C. Koster, M.~Kutschka, and
  U.~Steglich, ``Network planning under demand uncertainty with robust
  optimization,'' \emph{Communications Magazine, IEEE}, vol.~52, no.~2, pp.
  178--185, 2014.

\bibitem{LeguayNoms16}
J.~Leguay, L.~Maggi, M.~Draief, S.~Paris, and S.~Chouvardas, ``{Admission
  Control with Online Algorithms in SDN},'' in \emph{{Proc. NOMS}}, 2016.

\bibitem{BuchbinderN06}
N.~Buchbinder and J.~Naor, ``Improved bounds for online routing and packing via
  a primal-dual approach.'' in \emph{{Proc. FOCS}}, 2006.

\bibitem{Kesselheim2014}
T.~Kesselheim, A.~T\"{o}nnis, K.~Radke, and B.~V\"{o}cking, ``{Primal Beats
  Dual on Online Packing LPs in the Random-order Model},'' in \emph{{Proc. ACM
  STOC}}, 2014.

\bibitem{Agrawal2015}
S.~Agrawal and N.~R. Devanur, ``Fast algorithms for online stochastic convex
  programming,'' in \emph{{Proc. ACM SODA}}, 2015.

\bibitem{megiddo2003combine}
D.~P. de~Farias and N.~Megiddo, ``How to combine expert (and novice) advice
  when actions impact the environment?'' in \emph{{Proc. NIPS}}, 2003.

\bibitem{Wang08}
N.~Wang, K.~Ho, G.~Pavlou, and M.~Howarth, ``An overview of routing
  optimization for internet traffic engineering,'' \emph{Communications Surveys
  Tutorials, IEEE}, 2008.

\bibitem{ParisInfocom16}
S.~Paris, A.~Destounis, L.~Maggi, G.~Paschos, and J.~Leguay, ``{Controlling
  Flow Reconfigurations in SDN},'' in \emph{{Proc. Infocom}}, 2016.

\bibitem{yang2014traffic}
S.~Yang, F.~Kuipers \emph{et~al.}, ``Traffic uncertainty models in network
  planning,'' \emph{Communications Magazine, IEEE}, vol.~52, no.~2, pp.
  172--177, 2014.

\bibitem{malboubi2014intelligent}
M.~Malboubi, L.~Wang, C.-N. Chuah, and P.~Sharma, ``{Intelligent SDN based
  traffic (de) Aggregation and Measurement Paradigm (iSTAMP)},'' in
  \emph{{Proc. IEEE INFOCOM}}, 2014, pp. 934--942.

\bibitem{moshref2014dream}
M.~Moshref, M.~Yu, R.~Govindan, and A.~Vahdat, ``{DREAM: dynamic resource
  allocation for software-defined measurement},'' in \emph{Proc. ACM SIGCOMM},
  2014, pp. 419--430.

\bibitem{yu2013software}
M.~Yu, L.~Jose, and R.~Miao, ``Software defined traffic measurement with
  opensketch,'' in \emph{NSDI}, vol.~13, 2013, pp. 29--42.

\end{thebibliography}
\bibliographystyle{IEEEtran}

\end{document}